# COVID-19 Infection Analysis Framework using Novel Boosted CNNs and Radiological Images

Saddam Hussain Khan[1] *

[1]Department of Computer Systems Engineering, University of Engineering and Applied Science (UEAS), Swat, Pakistan;

**Email Address:** saddamhkhan@ueas.edu.pk

## Abstract

COVID-19 is a new pathogen that first appeared in the human population at the end of 2019, and it can lead to novel variants of pneumonia after infection. COVID-19 is a rapidly spreading infectious disease that infects humans faster. Therefore, efficient diagnostic systems may accurately identify infected patients and thus help control their spread. In this regard, a new two-stage analysis framework is developed to analyze minute irregularities of COVID-19 infection. A novel detection Convolutional Neural Network (CNN), STM-BRNet, is developed that incorporates the Split-Transform-Merge (STM) block and channel boosting (CB) to identify COVID-19 infected CT slices in the first stage. Each STM block extracts boundary and region-smoothing-specific features for COVID-19 infection detection. Moreover, the various boosted channels are obtained by introducing the new CB and Transfer Learning (TL) concept in STM blocks to capture small illumination and texture variations of COVID-19-specific images. The COVID-19 CTs are provided with new SA-CB-BRSeg segmentation CNN for delineating infection in images in the second stage. SA-CB-BRSeg methodically utilized smoothening and heterogeneous operations in the encoder and decoder to capture simultaneously COVID-19 specific patterns that are region homogeneity, texture variation, and boundaries. Additionally, the new CB concept is introduced in the decoder of SA-CB-BRSeg by combining additional channels using TL to learn the low contrast region. The proposed STM-BRNet and SA-CB-BRSeg yield considerable achievement in accuracy: 98.01 %, Recall: 98.12%, F-score: 98.11%, and Dice Similarity: 96.396%, IOU: 98.845 % for the COVID-19 infectious region, respectively. The proposed two-stage framework significantly increased performance compared to single-phase and other reported systems and reduced the burden on the radiologists.

# 1    Introduction

The new coronavirus (COVID-19) is a transmissible disease that first appeared in December 2019 and spread worldwide [1]. COVID-19 is an ongoing pandemic that has devastatingly affected the world [2]. The COVID-19 suspected cases are approximately 675 million, with 6.8 million deaths, while 647 million have been healthier. It is estimated that 99.6% of the infected patients have slight, while 0.4% have severe or critical symptoms [3]. However, it causes respiratory inflammation, difficulty breathing, pneumonia, alveolar damage, and respiratory failure in severe cases, eventually leading to death [4]. The person with COVID-19 pneumonia mostly depicts the signs of pleural effusion, ground-glass opacities, and consolidation [5].

The detection tests for COVID-19 include molecular testing (RT-PCR) and Chest radiological imaging (X-ray, CT scan) [6–8]. Chest imaging is also used to complement clinical evaluation, monitoring, and follow-up for COVID-19 diagnosed patients. Moreover, the CT scan is utilized for the severity assessment and treatment of COVID-19 patients. In a public health emergency, the manual examination of many radiological images is a great challenge and a severe concern for remote areas without experienced radiologists [9]. Radiological images usually are complex. The COVID-19 infected region has high variation in size, shape, and position. Furthermore, these radiological images are highly distorted due to noise during CT image acquisition [10]

Automatic detection technology is a serious need to help radiologists improve their performance and deal with many patients and will overcome the burden of manually examining. Therefore, Deep Learning (DL) based diagnostics techniques are developed to facilitate radiologists in identifying COVID-19 infection [11]. Such an effective predictive model can overcome the radiologist burden for manual assessment of COVID-19 infected CT, ultimately improving the survival rate. The contribution of DL and its capability to classify and segment the image with high accuracy will eliminate the probability of incorrect results by the currently used testing kits. DL will reduce the load on healthcare facilities [12,13].

The DL-based automated technique's remarkable success in different fields has attracted researchers to its application in medical diagnostic systems[14]. These tools are designed for automatic medical image analysis and facilitate radiologists in identifying lung-related anomalies [15]. These tools can detect minor irregularities of COVID-19 patterns that cannot be observed through a manual examination and reduce the burden on hospitals for COVID-19 diagnosis. COVID-19 specific radiographic patterns (region-homogeneity, textual variation, boundaries, etc.) are generally characterized by GGO, pleural effusion, consolidation, etc. [16]. Therefore, an integrated detection and analysis framework based on Convolutional Neural Network (CNN) is proposed that exploits COVID-19 specific radiographic patterns to identify and analyze COVID-19 infection in CT lung slices. The key contributions are fallow:

1. A deep CNN-based diagnostic system is developed for detecting COVID-19 infection and analyzing suspicious lesions in CT Lungs images to identify the severity and stage of the disease. The proposed system is categorized into two main phases: COVID-19 infection detection and infected region analysis using CT lungs images.
2. The new Split-Transform-Merge (STM) blocks and Channel Boosting (CB)-based deep STM-BRNet detection CNN is developed that extracts a diverse set of features to learn the COVID-19 specific patterns in CT images effectively. The STM block employed multi-path boundary and region-smoothing operations to learn homogeneous regions, texture variations, and boundary features.
3. A new deep SA-CB-BRSeg segmentation CNN is proposed to demarcate the COVID-19 infectious region in the CT lungs precisely. In this regard, average and max-pooling are employed systematically to exploit COVID-19 infection patterns related to region smoothing and discriminative features in encoder and decoder blocks.
4. A novel CB is introduced in the proposed decoder of SA-CB-BRSeg using transfer learning (TL) to learn low contrast infected regions. Moreover, the new attention block is implemented in the SA-CB-BRSeg segmentation CNN to effectively learn mildly infected regions.

The paper's layout is as follows: Section 2 provides COVID-19-related study. Section 3 describes the developed COVID-19 infection analysis framework, whereas Section 4 presents the material and implementation details. Section 5 discusses the results, and finally, section 6 concludes.

## 2	Related Works

Recently, CT technology has been used to diagnose COVID-19 infection in developed countries such as America, China, etc. However, manual examination of CT scans has a significant burden on radiologists and affects performance. Therefore, an analysis of the region of interest has been performed to detect the position and severity of the infection. Several classical techniques have been used for diagnosis but failed to show efficient performance [17]. Therefore, DL based tools are developed for quick infection analysis and facilitate the radiologist [18]. CNN is the branch of DL successfully used to analyze thoracic radiologic images to diagnose domain-specific COVID-19 features. A deep CNN-based model has been employed to automatically extract the most relevant dynamic features for COVID-19 infected patterns. These models can capture the infected region's useful dynamic features, discriminating the COVID-19 infected region from the healthy ones. This way, several deep CNNs, like VGG-16/19, ResNet-50, Xception, ShuffleNet, etc., have been employed on the COVID-19 CT dataset. These models achieved performance from an accuracy of 87% to 98% [15]. However, the aforementioned models have been employed for COVID-19 infection detection but need more analysis information.

Segmentation of contaminated regions, on the other hand, is commonly used to pinpoint the disease's location and severity. Some traditional segmentation approaches were used initially, but they could have delivered better results. Therefore, CNN-based VB-Net has been designed to segment COVID-19 infection in CT scans and stated a dice similarity (DS) of 91%. Moreover, the joint-classification-segmentation (JCS) framework has been reported to display and segment the contaminated zone using classification and segmentation[19]. This system achieved 95% sensitivity, 93 % specificity, and a DS score of (78.3%). Furthermore, the DCN method has been reported to analyze COVID-19 infection segmentation. The DCN method achieved an accuracy of 96.74% and a lesion segmentation DS of 83.51% [20] The COVID-19 diagnosis blocks corporates region and edge-based operations and collects diverse features. The technique achieved a 97% detection rate and 93% precision. The UNet and Feature Pyramid Network (FPN), with various encoder-backbone of DenseNet and ResNet, has been employed for lung region segmentation [21]. The reported technique showed a DS of 94.13% and IoU of 91.85%. The detection phase achieved a 99.64% detection rate and 98.72% specificity. Moreover, spatial and channel attention-based U-Net has been exploited to learn diverse contextual relationships to improve feature representation [22]. The obtained DS, detection rate, and Specificity are 83.1%, 86.7%, and 99.3%, respectively. Most of the previous works lack to address these challenges:

- The reported deep CNN techniques were trained on a small size of datasets. Evaluating the existing techniques on a small number of images limits the performance.
- Diagnosis is limited to detecting infected samples and, therefore, missing information about the disease's stages (mild, medium, and severe).

## 3    Methodology

This study developed a deep CNN-based detection and segmentation framework for automatically analyzing COVID-19-related abnormalities in the lungs. Diagnoses of the infectious regions are typically performed through segmentation to explore the location of the infection and disease severity [23,24]. The proposed system has three main technical innovations, (i) the proposed STM-BRNet detection model, (ii) The proposed SA-CB-BRSeg segmentation model, and (iii) the implementation of customized detection and segmentation CNN models. A new two-stage deep CNN-based diagnosis is developed for automatically analyzing COVID-19 irregularities in lungs. The proposed diagnostic system constitutes detection and segmentation phases. COVID-19 infected slices are separated from healthy individuals in the detection phase using CT images. While in the segmentation phase, the infectious region is segmented to identify the severity of the disease. The brief and detailed workflows of the proposed system are depicted in Figure 1.

### 3.1    COVID-19 Infection Detection

The proposed detection phase constitutes two modules (i) the proposed STM-BRNet detection CNN Model and (ii) Customized existing CNNs for comparative analysis. A new deep detection CNN is developed to discriminate COVID-19 infectious regions from healthy ones. The COVID-19 detection stage is illustrated in Figure 1.

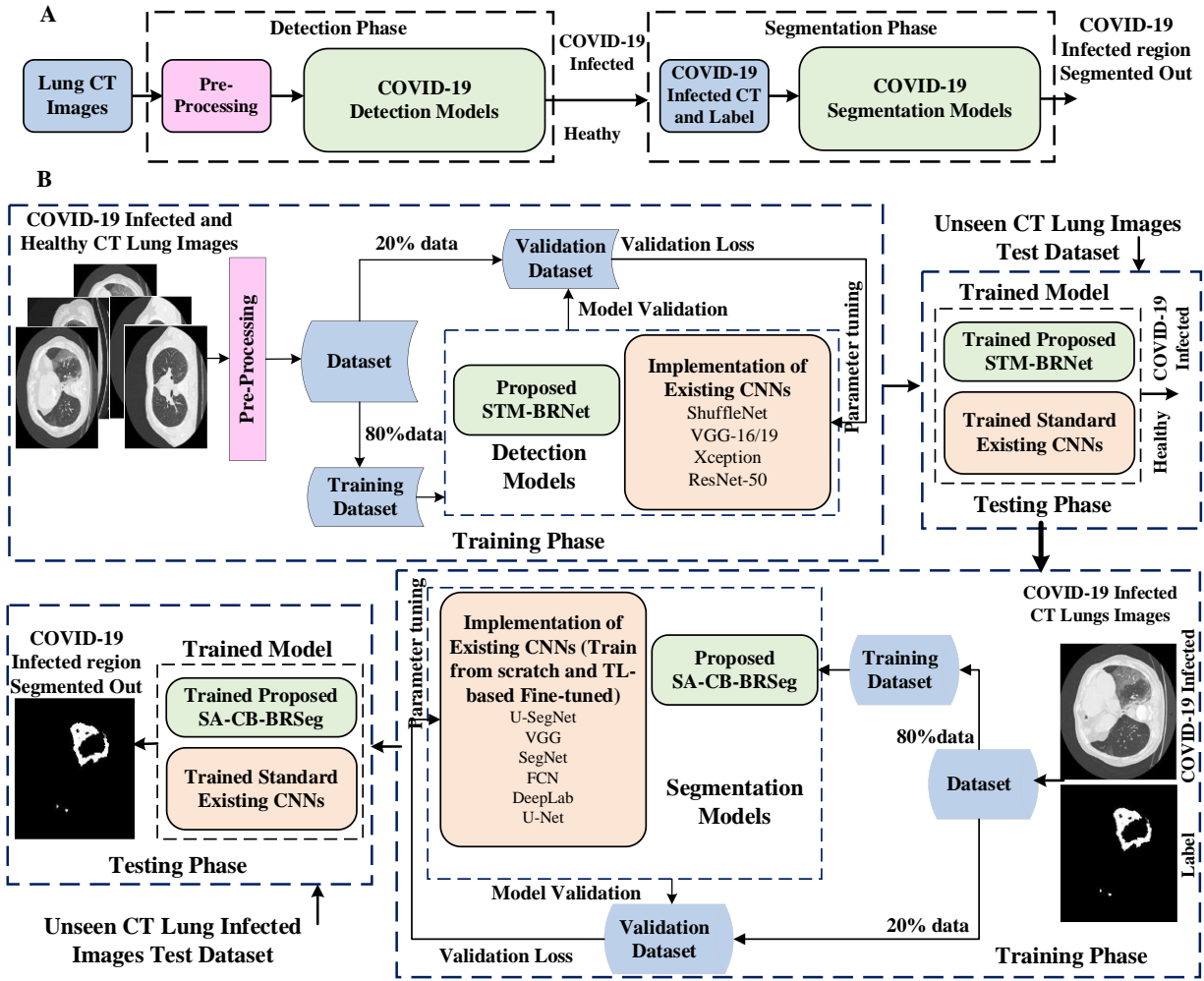

**FIGURE 1** Panel (A) shows the main steps for the proposed two-stage framework, whereas Panel B illustrates the complete workflow in detail

### 3.1.1 Proposed Detection STM-BRNet

This work develops a deep CNN, named "STM-BRNet", that effectively distinguishes COVID-19 infected images from healthy slices, shown in Figure 2. The significance of the proposed STM-BRNet is the systematic usage of new STM blocks and CB ideas. The proposed STM-BRNet CNN is based on the systematic use of dilated convolutions paired with the region and edge-based feature operations within STM blocks to learn the region-smoothing and structure of COVID-19 infected patterns. The STM-BRNet encompasses dilated convolutions that enhance the reception field and preserve data dimensions at the output layer to achieve a diverse feature set to distinguish infected regions from healthy areas [25]. Moreover, the new CB concept is altered at the STM blocks to preserve the reduced prominent maps and then joined to get various boosted channels and capture minor infection contrast variation. In addition, using various pooling operations results in down-sampling that eventually enhances the robustness of the model against any variation. Additionally, the region operator within the STM block utilizes the average pooling layer for smoothening and noise reduction.

**Architectural Design of the Proposed STM-BRNET**

The STM-BRNet comprises two STM blocks with identical topology and is arranged methodically to learn various features' initial and final levels. Each STM-RENet comprises four convolutional blocks, where Region and boundary operations are methodically employed. The dimension of each STM boosted block is 256 and 512 [26,27]. As the prime focus of the architecture is to get minor contrast and texture infection patterns, therefore, four diverse blocks, namely Region and Edge (RE), Edge and Region (ER), Edge (E), and Region (R), are implemented. The dilated convolutional layer, regional/boundary operations, and CB idea are altered to learn COVID-19 specific features in each block.

The RE block extracts regions and boundaries; it comprises two dilated convolutional layers followed by the average and max-pooling layers, as shown in equations (1-3). Moreover, the ER block extracts edges and regions; it comprises two dilated convolutional layers followed by a max-pooling layer. The E and R block learn the edges and smoothness, respectively. In block E, additional channels are generated by TL to achieve various channels, while block RE, ER, and E are learning from scratch.

These processes enhance the boundary information and region-specific properties, whereas dilated convolutional operations aid in learning the global receptive features. The perception of multipath-based STM blocks is used to obtain diversity in the feature set and can dynamically capture the minor representative and textural variations information from the COVID-19 infected CT images. Moroever, fully connected layers and dropout layers that preserve the prominent features and reduce overfitting.

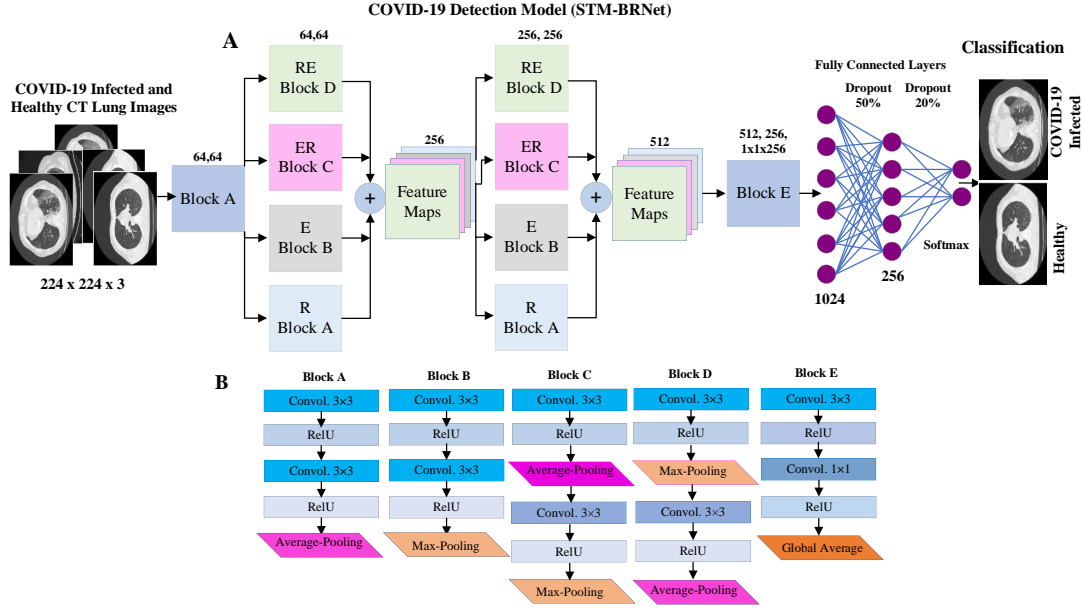

**FIGURE 2** Architectural design for the proposed STM-BRNet for COVID-19 Infection detection.

$$x_{k,l} = \sum_{i=1}^{m} \sum_{j=1}^{n} x_{k+i-1,l+n-1} f_{i,j} \tag{1}$$

$$x^{max}{}_{k,l} = max_{i=1,\dots,w, j=1,\dots,w} x_{k+i-1,l+j-1} \tag{2}$$

$$x^{avg}{}_{k,l} = \frac{1}{w^2} \sum_{i=1}^{w} \sum_{j=1}^{w} x_{k+i-1,l+j-1} \tag{3}$$

$$x_{Boosted} = b(x_{ER} || x_{RE} || x_R || x_E) \tag{4}$$

$$x = \sum_{a}^{A} \sum_{b}^{B} v_a \, x_{Boosted} \tag{5}$$

$$\sigma(x) = \frac{e^{x_i}}{\sum_{i=1}^{c} e^{x_c}} \tag{6}$$

The channels and size are represented by x and k x l. The kernels and their size are denoted by f and i x j in equation (1). In contrast, the output ranges to [1 to k-m+1, l-n+1]. Moreover, average and max-pooling window size is represented by **w,** respectively, on convolved output $(x_{k,l})$ (Equations 2-3). In equation (4), the original feature maps of block RE, ER, and R are signified by $x_{RE}$, $x_{ER}$, and $x_R$, respectively. Likewise, the auxiliary feature-maps of block R achieved using TL are denoted as $x_E$. These channels are boosted by concatenation operation b(.). The neuron quantity and activation in equation (6) are shown with $v_a$. and σ.

### 3.1.2 Implementation of existing Detection CNNs

Recently, CNN has demonstrated effective performance in medical field images to detect and segment medical images [14]. The employed models for detection are VGG-16/19, ResNet-50, ShuffleNet, Xception, etc. [28]. These deep CNNs with varying in-depth and network designs are tailored to detect and segment COVID-19 infected radiological images. Moreover, the initial and final layers are customized according to the target specific domain.

### 3.2 COVID-19 Infected Regions Segmentation

The proposed STM-BRNet aims to classify COVID-infected patients from healthy patients by utilizing the capabilities of deep CNN architectural ideas. The infected images are provided the segmentation CNNs for delineating COVID-19 infection regions that identify the disease's severity. This paper implements two different experimental setups for infection segmentation: (i) proposed SA-CB-RESeg segmentation, (ii) target-specific segmentation CNNs implementation from scratch, and TL.

### 3.2.1 Proposed SA-CB-RESeg Segmentation CNN

An SA-CB-RESeg is proposed to perform fine-grain pixel-wise segmentation. The proposed SA-CB-RESeg CNN is comprised of two encoders and boosted decoder blocks. The encoder and decoder blocks are designed in such a way as to improve the SA-CB-RESeg learning capacity. In this regard, average-pooling and max-pooling, along with convolutional operation in encoding and decoding stages, are employed systematically to learn region and boundary-related properties of COVID-19 infected regions [29,30]. Moreover, the convolutional operation employed a trained filter on images and generated feature maps of distinctive patterns. The encoders and decoders are designed symmetrically; however, in pooling operations, max-pooling is employed in the encoder for down-sampling. Contrarily, in the decoder, an un-pooling operation is employed to perform up-sampling. Finally, a 2x2 convolutional layer is employed to classify pixels into COVID-19 and background.

The encoder is designed to learn semantically meaningful COVID-19 specific patterns. However, the encoder loses spatial information essential for infected region segmentation because it reconstructs the infection map. In this regard, decoders are employed to preserve the spatial information of the corresponding encoders using pooling indices. These positional indices are stored in each pooling operation and are helpful for reconstruction and mapping on the decoder side. Moreover, the pooling operation performs down-sampling and reduces the spatial dimension (Figure 3).

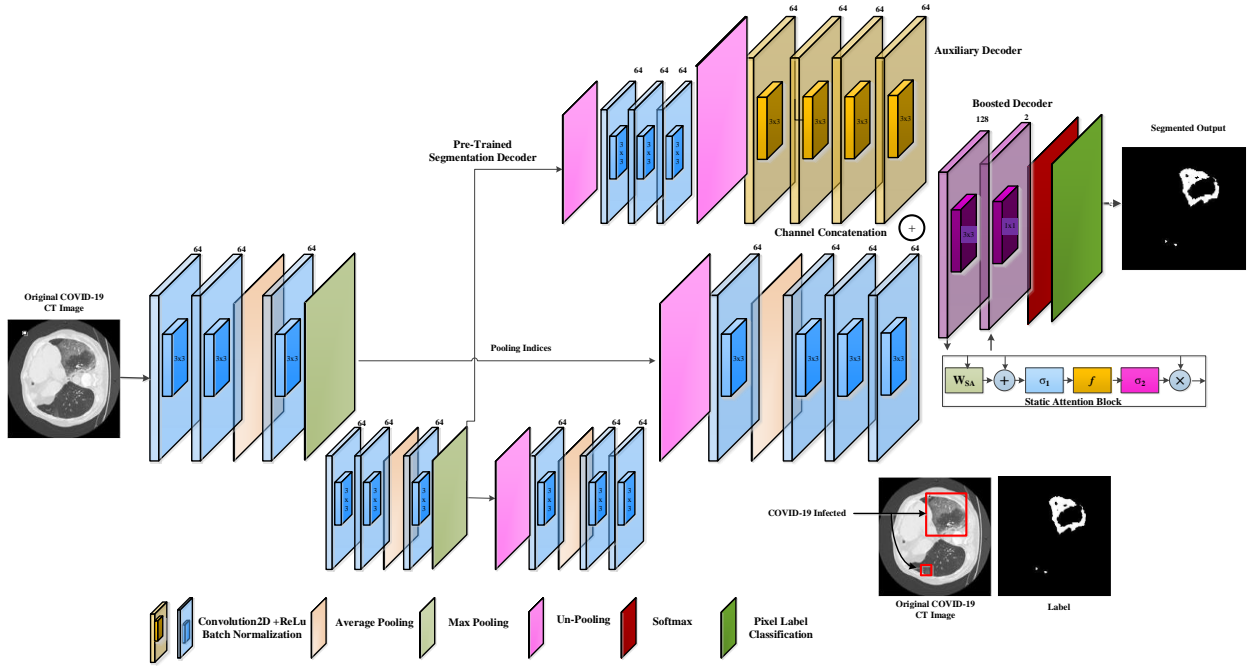

**FIGURE 3** Architectural design for the proposed SA-CB-RESeg

**Channel Boosting Significance**

The new CB idea is introduced by concatenation the original feature maps of the decoder with additional channels using TL to improve learning the low contrast COVID-19 infected region. The proposed SA-CB-RESeg utilized the additional channels generated from pre-trained CNN using TL combined with the original to get rich information feature maps and improve generalization. The SA-CB-RESeg benefited from learning from scratch and tuning on COVID-19 images using TL and CB. The boosting channels increase the SA-CB-BRSeg representative's capacity. Moreover, Finally, $X_{RE-e}$ and $X_{RE-d}$ are encoders (e) and decoders (d) blocks of the developed SA-CB-RESeg, as shown in Equations 7 & 8. Consequently, the boosting and auxiliary channel (AC) operation at the decoder side is shown illustrated in Equation 9.

$$X_{RE-e} = f_c(x^{avg} || x^{max}) \quad (7)$$

$$\mathbf{X}_{RE-d} = f_c(\mathbf{x}^{max} || \mathbf{x}^{avg}) \quad (8)$$

$$\mathbf{X}_{CB} = b(\mathbf{X}_{RE-d} || \mathbf{X}_{AC}) \quad (9)$$

**Static Attention**

Static attention (SA) enhances the learning capability of the COVID-19-infected areas by locating high weightage. The detail of the SA block is illustrated in Figure 4. $X_l$ demonstrates the input channel and $W_{pixel}$ is the pixel-weightage coefficient at the range of [0, 1] (Equation (10)). The output $X_{SA\_out}$ highlights the infected region while suppressing the irrelevant features. In Equations (11) and (12), $\sigma_1$ and $\sigma_2$ is the Relu and Sigmoid activation function, respectively. While $b_{SA}$ and $b_f$ is biasness, and $W_x$, $W_{SA}$, and $f$ is the linear transformation.

$$\mathbf{X}_{SA\_out} = W_{pixel} \cdot \mathbf{X}_l \quad (10)$$

$$X_{relu} = \sigma_1(W_x \mathbf{X}_l + W_{SA} SA_{m,n} + b_{SA}) \quad (11)$$

$$W_{pixel} = \sigma_2(f(X_{relu}) + b_f) \quad (12)$$

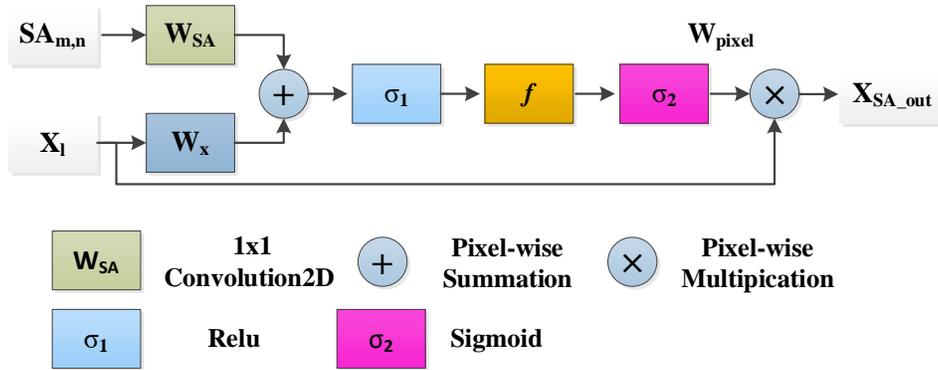

**FIGURE 4** Architectural design of Static Attention Block.

### 3.2.2 Implementation of Existing Segmentation CNNs

Several deep CNNs are employed to segment the COVID-19 CT infected region using diverse datasets [31]. This study employs implemented DeepLab, U-SegNet, SegNet, VGG-16, U-Net, and FCN as segmentation models [32–34]. The existing segmentation CNNs have been implemented for comparative studies. We have employed the existing CNN models by training from scratch and weight initialization. The weights are initialized from pre-trained CNNs using the concept of TL and fine-tuned on CT images.

## 4 Experimental setup

### 4.1 Dataset

Chest CT scan has a high sensitivity for the diagnosis of COVID. The major benefit of using lung CT scans is that it makes the internal anatomy more apparent as overlapping structures are eliminated, thus leading to efficient analysis of the affected areas in the lungs. To serve the purpose, a dataset of 30 patients with 1684 CT Lung images is used provided by the Italian Society of Radiology (SIRM) [35]. The dataset is based on CT lung images and their corresponding labels based on COVID-19 infected and healthy patients in .nii.gz format. The experienced radiologist examined the provided dataset. The CT samples were paired with a radiologist-provided binary label with marked infected lung regions.

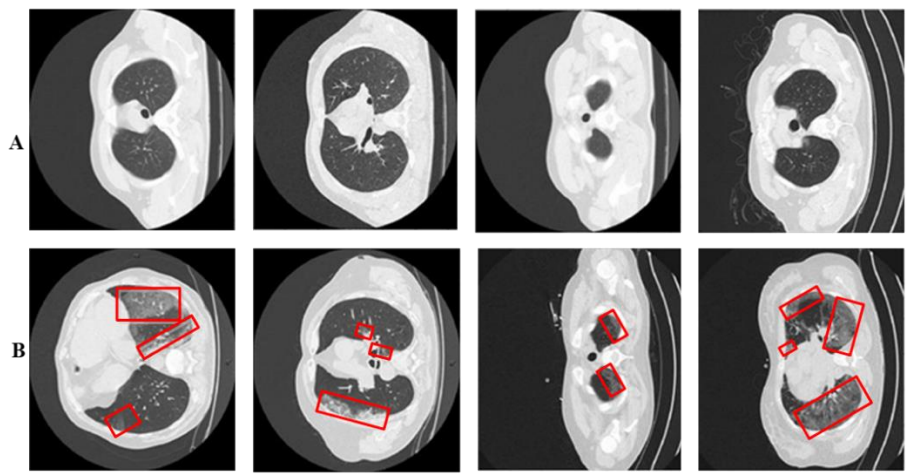

**FIGURE 5** Panel (A) and (B) represents COVID-19 infected vs. Healthy samples. While, the infected regions are red highlighted.

### 4.2 Implementation details

The detection and segmentation CNNs are trained separately in the developed diagnosis system. The dataset consists of CT lung images of COVID-19 infected and healthy patients. Thus we categorized the dataset into two classes, COVID-19 infected and healthy, depending upon their labels. The total number of images for the COVID-19 infected class is 852, and 832 for the healthy class. The CT lung images of COVID and healthy classes are used in the proposed architecture for classification purposes. For segmentation, only the COVID-19 infected class images (852) and their corresponding labels are used as it helps give a better insight for the analysis of the infected area.

The COVID-19 CT image datasets are divided into training and testing sets at the ratio of 8:2 for both the detection and segmentation phases. The training set is further divided into training and validation sets using hold-out cross-validation techniques. The proposed novel architecture based on Deep CNN detection and segmentation is built in MATLAB 2022a tool. The simulations are performed using an NVIDIA GTX-T HP computer and 64 GB RAM. Each model takes almost 13–23 hrs. ~1.5–2.5 hrs. /epoch, during training. The hyper-parameters regulate the optimization and convergence of the deep CNN models. Deep CNN models are trained for 10 epochs by selecting optimal hyper-parameters for smooth and efficient convergence [36]. The hyper-parameters detail is available in Table 1.

**TABLE 1** The selected hyper-parameters detail is used during training deep CNN for detection and segmentation.

| Hyperparameters | Values |
|---|---|
| Learning-rate | 0.0001 |
| Epoch | 10 |
| Model-Optimizer | SGD |
| Batch-size | 16 |
| Linear-Momentum | 0.95 |

## 4.3 Performance evaluation

The developed framework's performance is evaluated using standard measures, and its detail is illustrated in Table 2. The detection measures include accuracy, recall, etc., depicted in Equations (13-17). While the segmentation CNNs are assessed using IoU and DS coefficient that is expressed in Equations (18) and (19), respectively. Segmentation accuracy (S_Acc) to correctly predictction of positive and negative class samples. In comparison, S-Acc is used for the correct prediction of pixels. DS metric is used for structure similarity, and IoU is employed to identify the predicted vs. ground truth's overlapping ratio.

TABLE 2 Detail of detection and segmentation evaluation measures.

| Measure | Symbol | Detail |
|---|---|---|
| Accuracy | Acc | The ratio of detections that is correct out of all the predictions. |
| Recall | R | Correct predictions ratio (COVID-19 samples). |
| Specificity | S | Correct predictions ratio (Healthy samples). |
| Precision | P | Correct predictions ratio (COVID-19 samples) in overall predictions. |
| Mathew Correlation Coefficient | MCC | Identify the quality of confusion metrics on an unbalanced dataset. |
| Jaccard Coefficient | IoU | %The similarity between Label and predicted areas. |
| Dice-Similarity | DS | % The weighted_similarity between label and predicted areas. |
| Segmentation-Acc | S_Acc | %Pixels that are accurately partitioned into COVID-19 and Background. |

$$Acc = \frac{Correctly\ Predicted\ Slices}{Total\ Slices} \times 100 \tag{13}$$

$$P = \frac{Correctly\ Predicted\ COVID-19}{Correctly\ Predicted\ COVID-19 + Correctly\ Incorrectly\ Predicted\ COVID-19} \tag{14}$$

$$R = \frac{Accuratly\ Detected\ COVID-19}{Total\ COVID-19}S \tag{15}$$

$$S = \frac{Correctly\ Predicted\ Healthy}{Total\ Healthy} \tag{16}$$

$$F - Score = 2\frac{(P\ x\ R)}{P+R} \tag{17}$$

$$IoU = \frac{Correctly\ predicted\ infected\ region}{Correctly\ predicted\ infected\ region + Total\ infected\ region} \tag{18}$$

$$DS\ Score = \frac{2*Correctly\ Predicted\ infected\ region}{2*Correctly\ Predicted\ infected\ region + Total\ infected\ region} \tag{19}$$

## 5 Results

This paper proposes a new two-stage diagnosis framework to analyze the COVID-19 infectious region in the lungs. Distributing the proposed into two stages has two main advantages: improving the performance and reducing the computational complexities. Moreover, screening of COVID-19 infected samples and then analyzing the infectious region helps quickly identify the severity of the disease. Furthermore, the two-stage process rivals the clinical workflow, where patients are referred for further diagnostic tests after initial detection. The performance of the proposed STM-BRNet detection and SA-CB-RESeg segmentation CNNs are evaluated based on standard performance metrics. The proposed models are tested on unseen data and indicate considerable performance compared to existing CNNs.

## 5.1 Detection stage analysis

In this stage, a deep CNN-based STM-BRNet is developed to detect COVID-19-infected images. This stage is optimized with a high detection rate for recognizing the COVID-19 characteristic pattern and reducing false positives (Table 3). The learning ability of STM-BRNet for COVID-19 specific CT images is assessed and compared with existing CNNs. We optimized this stage for a high detection rate for recognizing the COVID-19 characteristic pattern with fewer false positives (shown in Table 3). The learning plot indicates accuracy and loss values for the training and validation dataset of the proposed detection STM-BRNet CNN (Figure 6). High training and validation error at the start has a maximum error; SGD fluctuates heavily. At the end of the training, SGD movement becomes smooth and reaches the solution.

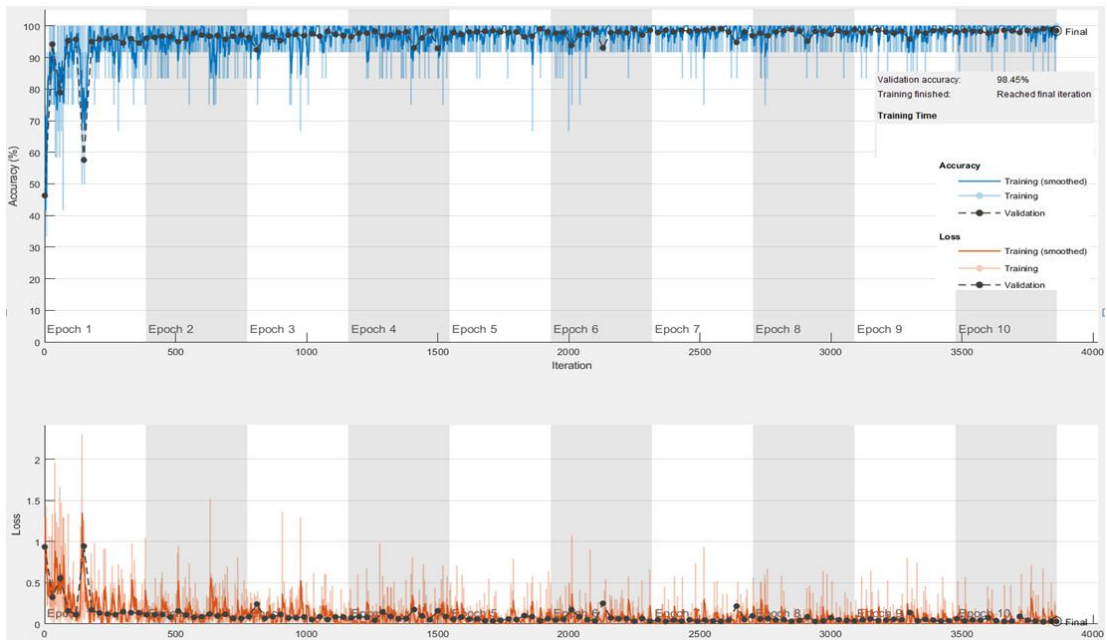

**FIGURE 6** Training convergence plot of the developed STM-BRNet detection CNN

### 5.1.1 The Proposed STM-BRNet's Performance Analysis

The proposed STM-BRNet is assessed on the test set using several performance measures like Accuracy, F-score, MCC, etc. The STM-BRNet achieves reasonable generalization compared to existing ResNet-50 in terms of accuracy (STM-BRNet: 98.01%, ResNet-50: 96.73%), F-score (STM-BRNet: 98.11%, ResNet-50: 96.77%), and MCC: (STM-BRNet: 94.85%, ResNet-50: 92.02%). The STM-BRNet technique of edge and region based STM blocks and CB using TL enhances the detection rate by classifying maximum samples as true positives. The high-intensity channel highlights the region boundaries, whereas the region approximation channel provides the region details. This fusion mimicked the idea of image sharpening using Laplacian of Gaussian to preserve optimal characteristics of infectious regions. Moreover, CB idea learned diverse feature-map from the pre-trained scenario and captured texture variations. These ideas methodical implementations improved the performance evident from Accuracy, MCC, and F-score (Table 3). The performance of the STM-BRNet is further increased by adding fully connected and dropout layers to emphasize the learning and improve the generalization.

**TABLE 3** Performance Comparison of the Proposed STM-BRNet and Existing CNNs.

| Model | Accuracy | F-score | ROC-AUC | PR-AUC | Precision | MCC | Specificity | Recall |
|---|---|---|---|---|---|---|---|---|
| ShuffleNet | 92.26 | 92.44 | 96.4 | 96.36 | 91.38 | 81.87 | 90.96 | 91.18 |
| VGG-19 | 94.35 | 94.43 | 98.33 | 98.84 | 94.15 | 87.21 | 93.98 | 92.94 |
| Xception | 95.83 | 95.81 | 99.09 | 98.58 | 97.56 | 90.67 | 96.99 | 91.67 |
| VGG-16 | 96.13 | 96.12 | 98.45 | 98.93 | 97.58 | 91.52 | 97.59 | 92.35 |
| ResNet-50 | 96.73 | 96.77 | 99.26 | 98.76 | 96.49 | 92.02 | 96.39 | 94.71 |
| **Proposed STM-BRNet** | **98.01** | **98.11** | **99.67** | **99.10** | **98.09** | **94.85** | **98.01** | **98.12** |
| **JCS** [19] | --- | --- | --- | --- | --- | --- | 93.00 | 95.00 |
| **VB-Net** [37] | --- | --- | --- | --- | --- | --- | 90.00 | 87.00 |
| **DCN** [20] | --- | 96.74 | --- | --- | --- | --- | --- | --- |

### 5.1.2 Performance analysis with the existing CNNs

The proposed STM-BRNet performance is compared with the five customized classification CNNs (VGG-16/19, ResNet-50, Xception, and ShuffleNet). The customized CNNs are famous for solving complex challenges and may be successively used to identify lung abnormalities. For a fair comparison, customized CNNs have been learned about the COVID-19 specific image. In contrast, the proposed STM-BRNet shows outperformance and performance gain in F-score, MCC, accuracy, etc., with the customized CNNs on the test dataset, as shown in Table 3 and Figure 9.

### 5.1.3 Features Visualization and PR/ROC Analysis

The considerable detection ability of STM-BRNet is evident from the principal components analysis (PCA) plot. PCA can be used to reduce the dimensionality of STM-BRNet features and identify the distinctive patterns for better discrimination. For comparison, deep feature-based analysis of best performing existing ResNet-50 is also provided in Figure 7. The considerable learning ability of the developed STM-BRNet is evident from the PCA plot that includes the first, second, and third generated principal components. Moreover, detection rate curves (PR/ROC) are also used to quantitatively assess the discrimination ability of the developed STM-BRNet (Figure 8). These are performance measurement curves that evaluate the generalization of the STM-BRNet by analyzing the discrimination between two COVID-19 infected and healthy classes at different threshold setups. Moreover, STM-BRNet has a good learning ability compared with different CNNs on the optimal threshold. The PR and ROC curve for COVID-19 detection based on deep STM-BRNet features gives a higher AUC, indicating better model performance.

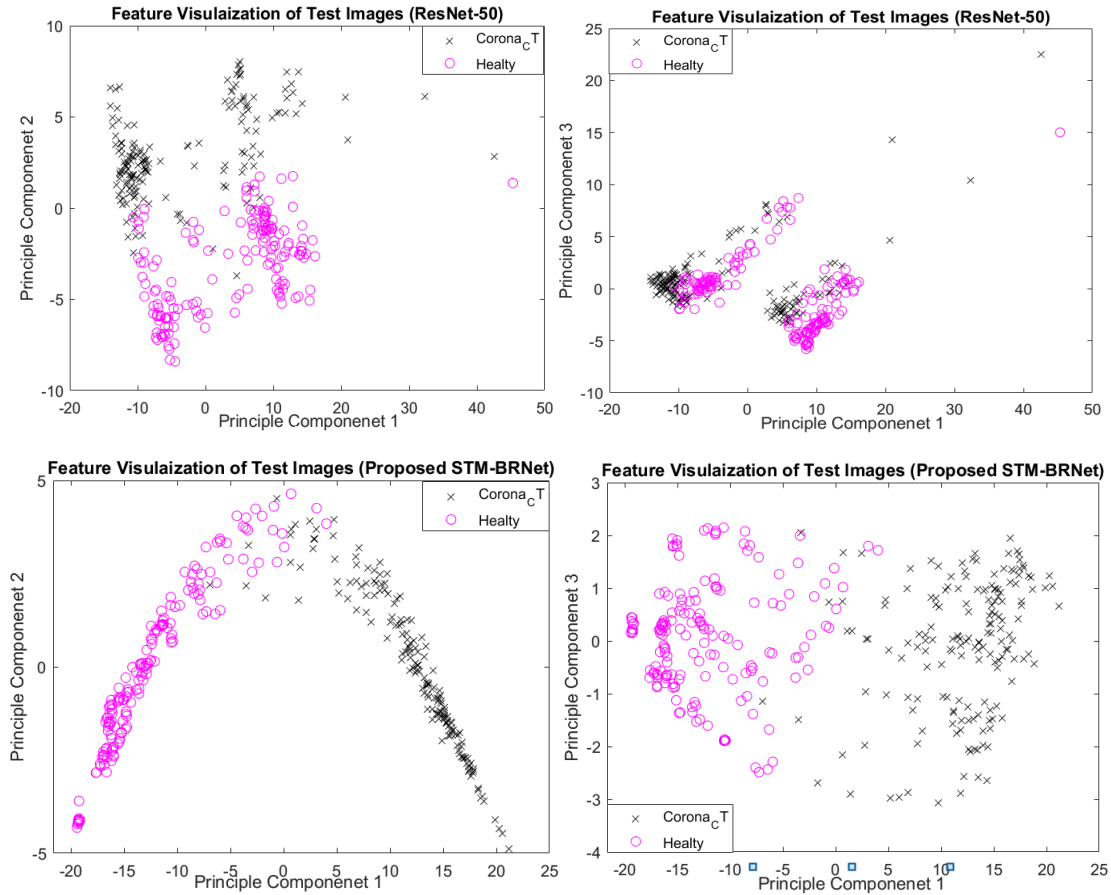

**FIGURE 7** Feature Space Visualization of the proposed STM-BRNet and ResNet-50 for the first, second, and third principal components generated.

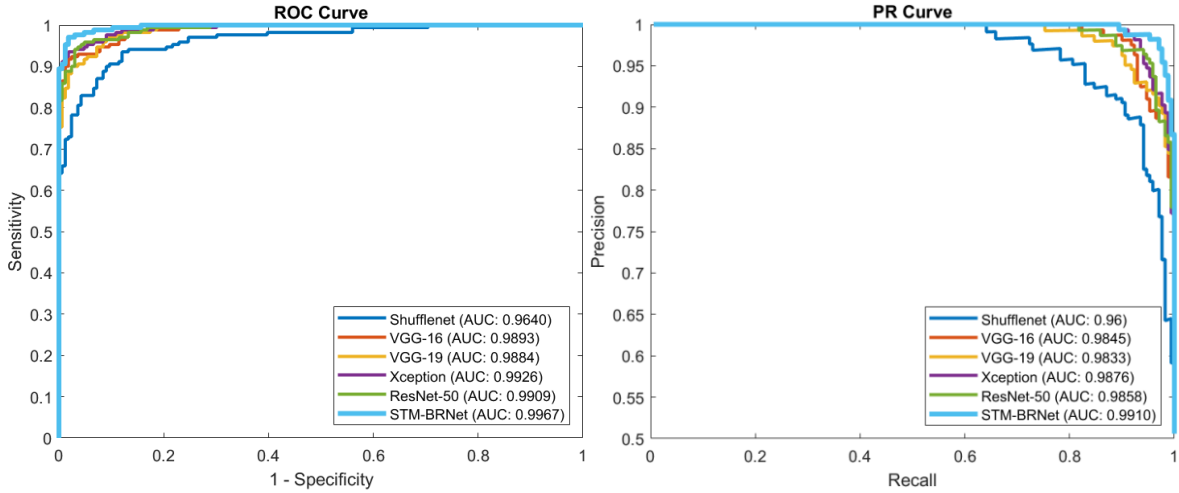

**FIGURE 8** PR and ROC curve for the developed STM-BRNet and customized existing CNNs.

## 5.2   Infected Region Analysis

CTs infected are separated using the developed STM-BRNet and assigned to deep segmentation CNN to analyze the infectious region. The infected slices and normal have minor contrast variations in the early stage. However, isolating the infected region from the healthy region is quite challenging. Therefore, the proposed SA-CB-RESeg segregates the infected regions by identifying infection boundaries and has minor contrast variation. Moreover, region analysis is needed to identify the severity of the mild, medium, or severe disease and its treatment design.

### 5.2.1.   Segmentation analysis of the proposed SA-CB-RESeg

The SA-CB-BRSeg is developed to segment COVID-19 infectious regions in CT lung images. The existing segmentation CNNs have been optimized based on COVID-19 infected specific patterns and imagery features. The experimental results on unseen data show the significance of the proposed SA-CB-BRSeg (Table 4). Moreover, the COVID-19 infected patterns vary in several patients. Furthermore, the pixel-wise segmentation of the infection region and created maps using the proposed SA-CB-BRSeg illustrate high subjective quality compared to existing CNNs (Figure 10).

In the proposed SA-CB-BRSeg, the new concept of region-homogeneity and boundary-based implementation is methodically incorporated using average and max-pooling. The systematic implementation and CB using TL help learn well-defined boundaries and texture variation in the infected lung region. Moreover, SA analysis of the infected region is required for an insight view of infectious patterns and their implication on surrounding or other organs. The results suggest that the proposed SA-CB-BRSeg has an excellent learning ability for COVID-19 infectious patterns, with evidence from DS and IoU scores of 96.40 % and 98.85 % (Table 4). In comparison, they precisely learned the discriminative boundaries and achieved a higher value of BFsa (99.09 %).

### 5.2.2. Segmentation Analysis with the Existing CNNs

The existing segmentation CNNs are employed to evaluate the learning capacity of the proposed SA-CB-BRSeg. In this regard, SA-CB-BRSeg performance is compared with six popular segmentation CNNs (DeepLabv3, U-SegNet, SegNet, U-Net, VGG-16, and FCN) (Table 4 and Figure 9). The segmented infected regions by the proposed SA-CB-BRSeg and existing CNNs are illustrated in Figure 10. The quantitative analysis recommends that the developed SA-CB-BRSeg performs than existing segmentation CNNs. However, the results show that customized CNNs perform poorly in learning mildly infectious regions. In contrast, VGG-16, FCN, and U-Net CNNs fluctuate in various stages of CT images and show less robustness in the models. The maximum accuracy in existing segmentation CNNs (Deeplabv3) is (98.48 %) for the infected region. Consequently, the DS score and IOU are (95 % against 96.40 %) and (97.59 % against 98.85 %), respectively.

The proposed SA-CB-BRSeg appears globally suited for moderate to severely infected regions. Moreover, the proposed and existing models' performance is improved using radiological and augmented data. The developed SA-CB-RESeg has low complexity and in-depth but shows more accurate performance than highly complex and large-depth models. Incorporating pixel-wise distribution of the developed SA-CB-RESeg improved the segmentation for various stages of infected regions. The performance metrics in Table 4 and the visual quality of the segmented maps in (Figure 10) evidence the outperformance of the proposed SA-CB-RESeg.

**TABLE 4** Performance analysis of the developed and existing segmentation CNNs (trained from scratch)

| CNN | Region | DS. | S-Acc. | IoU. | BFs. | Gl-Acc | Mn-Acc | Mn-IoU | Wt-IoU | Mn-BFs |
|---|---|---|---|---|---|---|---|---|---|---|
| **Proposed SA-CB-BRSeg** | **Infected** | **96.40** | **99.21** | **98.85** | **99.09** | **99.51** | **99.49** | **98.98** | **99.09** | **98.32** |
| | **Background** | **99.02** | **99.72** | **99.31** | **97.45** | | | | | |
| Deeplabv3 | Infected | 95.00 | 98.48 | 97.59 | 97.53 | 99.03 | 98.91 | 98.14 | 98.33 | 97.08 |
| | Background | 98.30 | 99.73 | 98.67 | 96.39 | | | | | |
| U-SegNet | Infected | 94.65 | 98.25 | 97.01 | 97.02 | 98.82 | 98.73 | 97.52 | 97.68 | 96.52 |
| | Background | 98.01 | 99.16 | 98.10 | 95.22 | | | | | |
| SegNet | Infected | 94.30 | 98.97 | 96.56 | 96.73 | 98.72 | 98.71 | 97.18 | 97.32 | 96.48 |
| | Background | 97.90 | 98.45 | 97.79 | 95.22 | | | | | |
| U-Net | Infected | 94.00 | 98.61 | 95.98 | 96.91 | 98.40 | 98.44 | 96.70 | 96.87 | 95.49 |
| | Background | 97.70 | 98.28 | 97.42 | 94.07 | | | | | |
| VGG-16 | Infected | 91.00 | 91.38 | 88.91 | 89.37 | 95.59 | 94.81 | 91.05 | 91.53 | 83.66 |
| | Background | 95.00 | 98.25 | 93.18 | 77.95 | | | | | |
| FCN-8 | Infected | 90.70 | 90.92 | 89.11 | 87.74 | 95.32 | 94.55 | 90.63 | 90.20 | 82.43 |
| | Background | 94.00 | 98.18 | 92.15 | 77.11 | | | | | |
| VB-Net [37] | Infected | 91.00 | --- | --- | --- | --- | --- | --- | --- | --- |
| DCN [20] | Infected | 83.50 | --- | --- | --- | --- | --- | --- | --- | --- |
| JCS [19] | Infected | 78.50 | --- | --- | --- | --- | --- | --- | --- | --- |

Gl-Acc, Mn-Acc. represents global and mean accuracy where Mn-IoU and Wt-IoU denote means and weighted IoU.

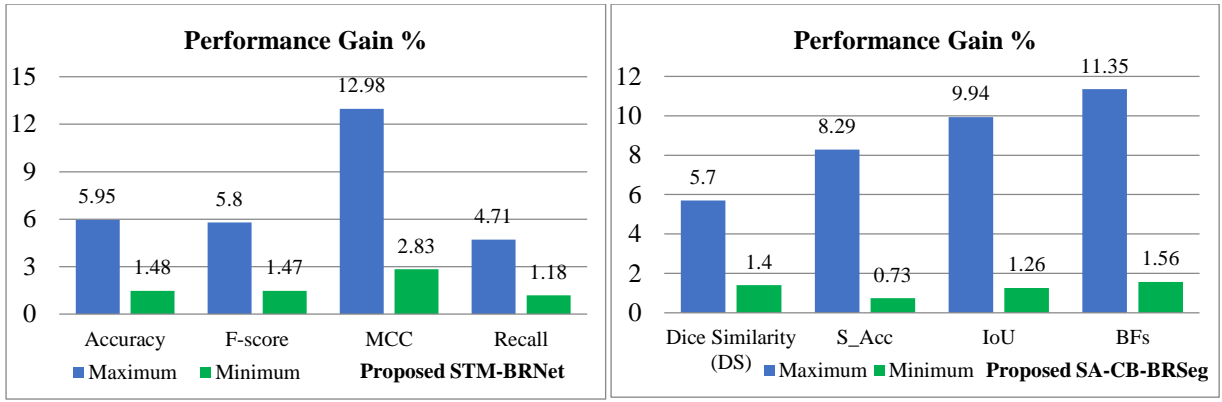

**FIGURE 9** The proposed STM-BRNet and SA-CB-BRSeg performance gain over the existing CNNs.

### 5.2.3. TL based Segmentation Analysis

TL-based fine-tuned also improved the performance gain over training from scratch in terms of DS score (0.35 to 4%), S-Acc (0.28% to 7.37 %), DS score (0.44 to 8.13%), and BFs (0.66 to 9.34%), as shown in Tables 4-5. Performance gain suggests that TL enhances performance as compared to learning from scratch. In this regard, TL–based feature maps are generated and concatenated to the decoder of the proposed SA-CB-RESeg. TL enhances the model convergence and generalization by achieving optimized weights or learned patterns from pre-trained scenarios [38,39]. Moreover, the radiologist labeled and augmented data are combined to improve the proposed SA-CB-RESeg performance[40]. The segmentation for the analysis of infected regions is achieved using the best-performing existing TL-based trained DeepLabv3 CNN for comparative analysis. The DeepLabv3 has gained accuracy (98.76%) and IOU (98.03%) for COVID-19 infected regions (Table 5).

**TABLE 5** Performance of segmentation CNNs (Implemented using TL).

| CNN | Region | DS. | S-Acc. | IoU. | BFs. | GL-Acc | Mn-Acc | Mn-IoU | WT-IoU | Mn-BFs |
|---|---|---|---|---|---|---|---|---|---|---|
| Deeplabv3 | Infected | 95.35 | 98.76 | 98.03 | 97.71 | 99.23 | 99.15 | 98.40 | 98.48 | 97.19 |
| | Background | 98.65 | 99.53 | 98.76 | 96.66 | | | | | |
| U-SegNet | Infected | 95.20 | 98.41 | 97.52 | 97.46 | 98.93 | 98.81 | 98.10 | 98.25 | 96.94 |
| | Background | 98.60 | 99.67 | 98.61 | 96.32 | | | | | |
| SegNet | Infected | 95.10 | 98.29 | 97.70 | 97.41 | 99.10 | 98.95 | 98.09 | 98.22 | 96.82 |
| | Background | 98.10 | 99.62 | 98.55 | 96.23 | | | | | |
| U-Net | Infected | 94.90 | 98.74 | 97.62 | 98.19 | 99.22 | 99.06 | 98.06 | 98.15 | 96.63 |
| | Background | 98.20 | 99.07 | 98.49 | 95.86 | | | | | |
| VGG-16 | Infected | 94.80 | 98.29 | 97.07 | 97.11 | 98.89 | 98.79 | 97.61 | 97.74 | 96.18 |
| | Background | 97.90 | 99.23 | 98.24 | 95.29 | | | | | |
| FCN-8 | Infected | 94.70 | 98.29 | 97.04 | 97.08 | 98.87 | 98.76 | 97.59 | 97.71 | 96.16 |
| | Background | 98.00 | 99.21 | 98.15 | 95.27 | | | | | |

### 5.2.4. Visual Analysis of the Proposed SA-CB-BRSeg

Visual analysis of COVID-19 infection segmentation using deep SA-CB-BRSeg is used to identify and analyze infected regions. The subjective evaluation shows that the proposed SA-CB-BRSeg accurately highlights the infected region. Incorporating pixel-wise distribution of the proposed SA-CB-BRSeg improved the segmentation of various stages of infected areas. The performance metrics in Table 5 and the visual quality of the segmented maps in (Figures 10 & 11) are evidence of the outperformance of the proposed SA-CB-BRSeg. Our proposed detection and segmentation CNNs are thoroughly trained and ready to be tested on unseen images. Finally, the existing segmentation CNNs, trained from scratch and TL-based, are also analyzed.

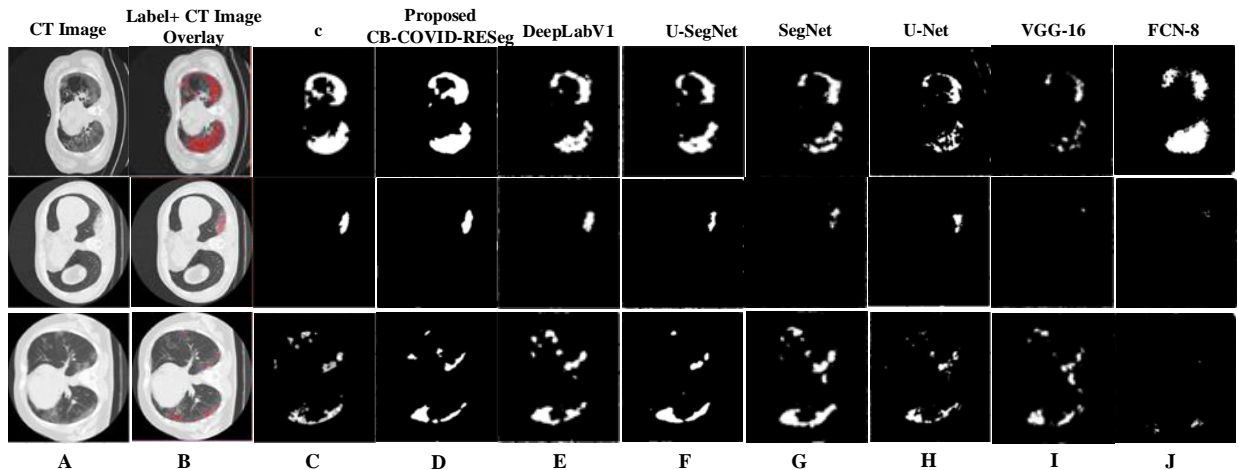

**FIGURE 10** Visual Evaluation of the proposed SA-CB-BRSeg and existing segmentation CNNs results.

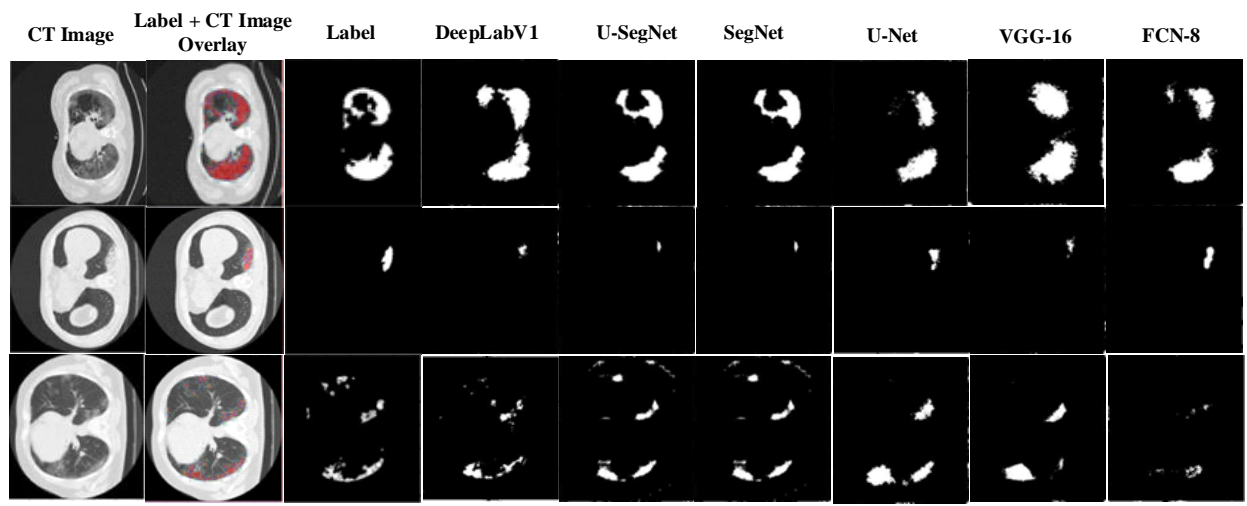

**FIGURE 11** Visual Evaluation of TL-based segmentation CNNs results on the test set.

## 6. Conclusions

COVID-19 is a transmissible disease that has primarily affected worldwide. These CT Lungs images exhibit distinctive patterns associated with COVID-19 abnormalities. Early diagnosis of COVID-19 infection is required to control the spread of the disease. In this research, a new deep CNN-based two-stage diagnosis is developed to detect and analyze COVID-19 infectious regions. This integrated approach exploits diverse features, including homogeneous regions, texture variations, and boundaries, to effectively learn the COVID-19 patterns. The proposed STM-BRNet benefits from data augmentation, TL-based diverse maps generation, and STM blocks. The significant discrimination ability of STM-BRNet screening model (F-score (98.11%), accuracy (98.01%), and recall (98.12%)) as compared to the existing deep CNNs on the test dataset. We have also demonstrated through simulations that the proposed SA-CB-BRSeg (IoU: 98.85%, DS: 96.40%) can precisely identify and analyze the infectious regions in CT. The proposed SA-CB-RESeg benefited from training from scratch and fine-tuning on COVID-19 data using TL and CB. The integrated approach discovers the whole COVID-19 infected region, which may help the radiologist estimate the disease's mild, medium, and severe stages. In contrast, a single-phase framework may not effectively give a precise and accurate detailed analysis of the infected region. COVID-19 is a novel infectious disease, and publically available labeled samples are limited. Therefore, in the future, we will employ the proposed framework on big datasets to improve the reliability of real-time diagnostics. Moreover, the dataset can be increased using the augmentation of the training sets by generating synthetic examples using GAN. Furthermore, it may be modified to segregate the infectious region into multi-class characteristic patterns automatically.

## Acknowledgments

We also thank the Department of Computer Systems Engineering, University of Engineering and Applied Sciences (UEAS), Swat, for providing the necessary computational resources and a healthy research environment.

## Declarations

### Conflicts of interest
Authors declared no conflict of interest.

### Availability of data and material
Publicly available dataset is used in this work that is accessible at
https://medicalsegmentation.com/covid19/
https://zenodo.org/record/3757476#.YVyHNdpBw2w